\documentclass[aps,floats,pre,graphicx,twocolumn,showpacs,preprintnumbers,amsmath,amssymb,french]{revtex4} 
 \usepackage{graphicx}
 \usepackage{dcolumn}
 \usepackage{bm}
 
 \newcommand{\gttwc}{\ensuremath{|g_1(t_w,t)|^2}}
 \newcommand{\gttwg}[1]{\ensuremath{|g_1(t_w,t,\gamma=#1 \%)|^2}}
 \newcommand{\dr}{\ensuremath{P(\tau_{rel})}}
 
\begin{document} 
 \title{Aging and effective delays in colloidal glasses under shear} 
 \author{V.Viasnoff, F.Lequeux } 
 \affiliation{ L.P.M UMR 7615 CNRS ESPCI 10 rue Vauquelin 75231 Paris, France } 
 \begin{abstract} 
 In this paper we analyze the changes in the microscopic dynamics of a colloidal glass submitted to an oscillatory shear. We use Multispeckle diffusing Wave Spectroscopy to monitor the transient dynamical regimes following a shear application. We show that the system displays a spontaneous aging that is amplified by a low amplitude oscillation (overaging) but stopped by a high amplitude one (rejuvenation). Intermediate amplitudes drive the system into a dynamical state that cannot be reached through spontaneous evolution. We demonstrate that theses systems present a weak long term memory and that the all observed phenomenology can be explained by a broadening of the relaxation time distribution under oscillatory strain. We discussed the similarities of our results with the behavior of spin glasses upon temperature shifts. We eventually propose a scenario based on the existence of harder and softer regions in the material.
 \end{abstract} 
    \date{\today}
    \maketitle
  
 \section{Introduction}
 It has been known for long that many physical systems such as molecular liquids, metallic liquids, polymer melts when cooled at low temperature exhibit a dynamical transition without structural changes. Their typical relaxation time increases by more than 10 decades and can no longer be measured when the temperature is low enough.
This phenomenon is known as the glass transition. When the typical relaxation times becomes superior to any reasonable experimental time scale, the systems displays a non-equilibrium behavior. This out-of-equilibrium behavior is revealed by a continuous slowing down of the microscopic dynamics of the system. This phenomena is referred as \textit{aging}.
\par Aging processes have been extensively studied in the past decades. The basic experiments to test aging consist in quenching  the system from the high temperature phase into the glassy phase and then measuring the time evolution of its physical properties such as response or correlation functions. The non equilibrium behaviour appears in the dependance of theses functions with the time $t_w$ that the system had spend in the glassy phase. For these simple thermal histories $t_w$ is called the age of the system. The slowing down of the dynamics following such a rapid quench is coined as simple aging (see eg:\cite{ Struikbook,Vincent96}). It often happens in a self similar manner such that the responses or correlation functions can  be rescaled on a master curve when plotted against the reduced variable $\frac{t}{t_w^\mu}$. $\mu$ is then called the aging exponent\cite{Bouch97}. This past decades a great number of experiments showed the universality of this kind of behavior for an increasing variety of systems\cite{Kityk02,AngellRev00,Kircher02}. \\
The glassy state can also be reached by increasing the density. This can be realized on colloidal glasses \cite{VanMeg94} that also exhibit a glass transition upon an increase of volume fraction \cite{Knaebel00,Ramos01}. When rapidly driven into their glassy region, this systems also display classical aging with an aging exponent usually inferior to 1. However temperature and density are not practical control parameters for colloidal glasses. It has nonetheless been observed \cite{Derec00,Cloitre00} that theses type of systems can be fluidized under mechanical sollicitation (oscillatory shear strain or constant shear rate). Moreover, it has been recently shown that the slowing down of the dynamics following the stop of the mechanical drive is that of a simple aging experiment. Hence the cessation of high shear is often referred as a rheological quench.
\par Simple aging experiments are the simplest protocoles that can be implemented on glassy systems. However, more elaborate temperature histories gave deeper insight of the organisation of the microscopic dynamics of theses amorphous materials. A common temperature history that has been tested on this systems is a transient temperature shift. After a quench into the glassy phase, the system is transiently driven from a non equilibrium state into another non equilibrium state. This protocole implemented on spin \cite{Dupuis01}, molecular \cite{Kityk02}, polymer \cite{Bellon02}, ferroelectric \cite{Doussineau02} glasses cast into light new phenomena such as rejuvenation, memory effects or Kovacs effects \cite{Kovacs79}. The two first effects revealed  a very peculiar organisation of the phase space of theses systems (hierarchical organization \cite{Lederman91} or chaotic behavior \cite{Fisher86,Bray87}). The last effect has been recently interpreted as a non monotonic change of the dynamics at all length scales \cite{BertBouch02}. In this paper we transpose this more elaborate temperature protocole to a colloidal glass using transient mechanical shearing periods instead of transient temperature shifts. The observed phenomenology following this two types of perturbations can thus be compared.
\par We use MultiSpeckle Diffusing Wave Spectroscopy  (MSDWS) to probe the transient change of the microscopic dynamics upon shear application. In the  first part of this paper we will recall the technique. We will secondly confirm that colloidal glasses can be driven into a steady reproducible state by high oscillatory shear. We also found that the systems has a full aging behavior after the shear cessation. We then show that the system can keep aging under low enough shear. We thus determine the shear amplitude that drive the system into a steady regime. Thirdly, we show the change of the dynamics after a transient mechanical perturbation can be described by three regimes. High amplitudes perturbations lead to a rejuvenation of the system whereas low amplitude ones help the system to age (overaging). Intermediate amplitudes lead to a dynamical state where both signs of overaging and rejuvenation are simultaneously present depending  on the considered time scales. We discuss a description of the two first regimes in terms of effective delay. We then explain why effective delay is an insufficient notion to describe our result and propose a model independent interpretation of our data in terms of changes of shape of the relaxation time distribution for the system. We show that our results  are consistent with the existence of a weak long term memory and that they do not give any hint of a hierarchical organisation of the relaxation times. We eventually propose a spatial scenario to explain the observed phenomenology.
\section{Experimental methods}
In this section we describe the sample we used as well as the MSDWS technique.
    \subsection{Sample preparation}
    The sample is a commercial suspension of polystyrene spherical beads of diameter 162 nm copolymerized with acrylic acid (1\%) that creates a charged corona stabilizing the microspheres. The corona prevents from aggregation by steric and electrostatic repulsions. A concentrated suspension of theses colloidal particules is prepared by dialysis from a dilute suspension. Great care is taken that no air bubble is introduced in the dense suspension during sample preparation. The volume fraction of the sample is fixed at $\phi=50\%\pm 0.5\%$ and was determined by drying. For that volume fraction the sample is pasty and displays a weak yield stress. No sign of crystallization could be noticed over several month. The volume fraction was chosen to fulfill two criteria:
\begin{itemize}
  \item The sample has to be concentrated enough so that the structural relaxation can be observed ($\alpha$ modes). 
  The $\alpha$ modes must also display a measurable aging behavior over all the experiment time scale.
  \item The volume fraction has to be low enough so that the strain is homogeneous, at least on macroscopic length scale all through the sample. We used a high resolution camera and a x20 zoom to check the shear strain was homogeneous at a macroscopic level over the whole sample. Effect like shear banding or wall slippage were not detected.
\end{itemize}
The sample is then carefully loaded through a syringe into a custom made shear cell. The cell is sealed with silicon oil which viscosity is chosen so that it forms a stable meniscus around the cell for several months.

    \subsection{Diffusing wave spectroscopy set-up}
\begin{figure}[th]
\begin{center}
\includegraphics[width=8cm,clip]{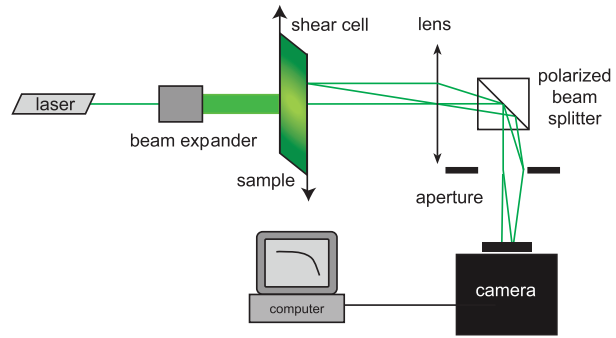}
\end{center}
\caption{Experimental set-up. The Argon-ion laser is first expanded. The beam is then sent onto the shear cell. The emerging light is collected by a lens and focus onto the camera CCD chip through an iris. The speckle pattern is analyzed in real time via a custom-made software. }
\label{dessinmontage}
\end{figure}

    Since we extensively described the Multispeckle Diffusing Wave Spectroscopy (MSDWS) in a previous article \cite{ Viasnoffrsci02}, we will simply recall the principle of this technique as well as our experimental set-up.\\
MSDWS is a dynamical light scattering technique that enables the monitoring of particles motions in a turbid non ergodic media. We use an Argon-ion laser (\textit{Spectra Physics BeamLok} 2Watts). The beam is expanded to a diameter of 1 cm and is incident on the plate-plate shear cell with a 3mm gap.  Multiply scattered light is
collected by a 50-mm Nikon camera lens. It is focused onto an iris diaphragm.
The lens is set up so that a one-to-one image of the scattered light from
the output plane of the sample cell is formed on the diaphragm. The scattered light emerging from the diaphragm is detected by a Dalsa CCD camera, model CAD1-128A,
which is placed approximately 15 cm behind the diaphragm. It is an
8-bit camera which we run at 500 frames per second. The images are
transferred to a computer by a National Instruments
data acquisition board, model PCI1422. Figure \ref{dessinmontage} shows the experimental set-up.
The interference state of the light collected onto the CCD chip is a random pattern called a speckle pattern. We analyse the temporal fluctuations of this pattern via the two times intensity autocorrelation :
$$ g_2(t_w,t)=\frac{\left\langle I_i(t_w)I_i(t+t_w)\right\rangle_i}{\left\langle I_i(t_w)\right\rangle_i \left\langle I_i(t+t_w)\right\rangle_i}$$
where $I_i(t_w)$ is the intensity recorded on pixel $i$ at time $t_w$, and $\left\langle...\right\rangle_i$ is the average over all the pixels of the considered frames. The intensity autocorrelation function can be related to the normalized electric field autocorrelation function $g_1(t_w,t)$ via the Siegert relation:
$$\gttwc=\frac{1}{\beta}( 1-g_2(t_w,t))$$
where $\beta $ is a normalizing factor.
 This technique does not require any time average. It thus allows a precise study of non ergodic systems in transient regimes (see ref \cite{Viasnoffrsci02}). Although the direct extraction of the scatters motion from \gttwc is far from being established in the case of dense sample, we will show that interesting informations can be extracted from the changes of correlation function shape. 
\section{Experimental results}
In this section, we present the different dynamical behavior displayed by our colloidal glass when submitted to an oscillatory shear. We first show that a complete rejuvenation of the system can be achieved under a high shear strain. After the shear cessation the colloidal glass displays a generic aging behavior. We then demonstrate that the aging process survives to a low amplitude shear. Under such conditions, not only the system keeps on aging but the aging rate is made faster by the shear application. The dynamics thus appears older. We call this phenomenon: "overaging". For intermediate shear amplitude, the system presents both behavior depending on the different probed time scales.

    \subsection{Complete rejuvenation and simple aging}\label{cr}
\begin{figure}[th]
\begin{center}
\includegraphics[width=8cm,clip]{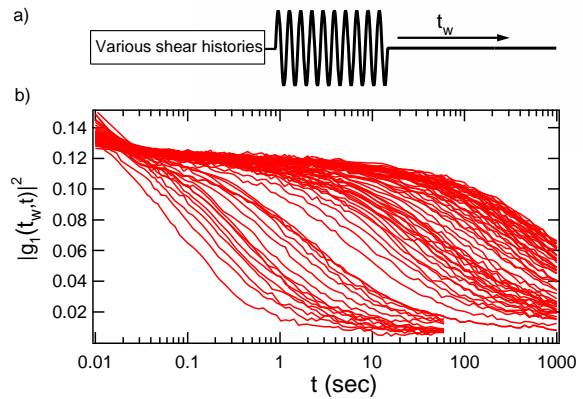}
\end{center}
\caption{a) Sketch of the shear history undergone by the sample. b)Intensity autocorrelation function \gttwc measured in backscattering. From left to right the different curves correspond to increasing $t_w$ (from 0.1s to 1500s). Notice that the typical decorrelation time increases with $t_w$. The shape of the correlation function is invariant on a log scale. This behavior is typical of an aging behavior.}
\label{vieilsimple}
\end{figure}

        We first establish that the microscopic dynamics of the system can be driven to a steady state by the application of a high shear. Whatever the past history of the sample, its microscopic dynamic can resume an equilibrium behaviour in a finite amount of time if the sample undergo a mechanical shearing of high amplitude. Indeed, as presented on figure \ref{vieilsimple} (a), we submitted our sample to various histories. They consist in a series of both different periods of shearing at various rates and amplitudes and of periods of rest for various duration. The sample is then submitted to an oscillatory shear of frequency f=1Hz and amplitude A=30\% during D=100s. After the shear cessation the sample is left at rest and its microscopic dynamics is probed by measuring the two times intensity autocorrelation functions $\gttwc$. The results are presented on figure \ref{vieilsimple} (b).

 Whatever the history before the oscillations may be, the set of curves obtained after the shear cessation is always the same. Further more these curves remain unchanged, within our experimental precision, when the duration D varies from 40 sec to 10000 sec, and/or when the amplitude $A$ ranges within 20\% and 35\% (that corresponds to the upper limit of our shear cell). Hence applying an oscillatory strain for 100s at 30\% erases all the memory of the past history of the sample. It is then completely rejuvenated. As demonstrated on figure \ref{vieilsimple}(b), the system starts aging when the shear is stopped. The auto correlation functions depend both on t, the time elapsed since the beginning of the measurement, and $t_w$, the time elapsed since the shear cessation. All theses curves can be rescaled and their characteristic decorrelation times $\tau_\alpha$ (define as the times needed for the functions to decorrelate by half their initial values) vary like $t_w^{1.06\pm 0.08}$ as demonstrated in \cite{viasnoffprl02}. The cessation of the high shear can thus be view as a rheological quench.
    \par
\begin{figure}[th]
\begin{center}
\includegraphics[width=8cm,clip]{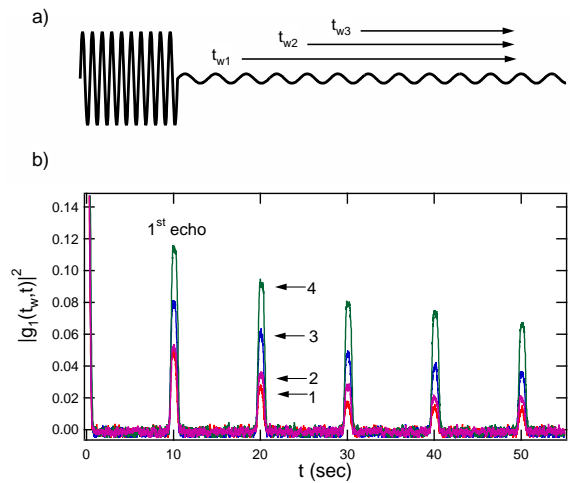}
\end{center}
\caption{a) Sketch of the shear history undergone by the sample. The waiting time $t_{w_i}$ at which the correlation functions are started coincide with a maximum of the shear amplitude. b)Intensity autocorrelation function \gttwc measured while the sample is sheared at 5.5\% at a frequency of 0.1Hz. The curves are labelled 1,2,3,4 from bottom to top. The waiting time $t_{w_1}$ are respectively 10,60,100,300s. The curves present an echo at every period. The height of each echo increases with $t_{w_i}$ up to 1500s, the longest measured waiting time (not represented for sake of clarity). The system does not reach any steady state on our measurement time but keeps on aging under shear.}
\label{dynaforce}
\end{figure}

 The system is now submitted to a shear strain of amplitude 30\%, 1Hz for 100s. The amplitude is then lowered to 3\%, the frequency is changed to 0.1Hz and the correlation functions are recorded while the shear is \textbf{On}. This history is sketched on figure \ref{dynaforce}(a). Whereas previously, the decorrelation was  due only to thermally driven motions of the scatterers, we now monitor together thermal and forced motions of the latex spheres due to the external shear. The characteristic shape of $\gttwc$ is given by fig \ref{dynaforce}(b). The curves  displays a pseudo-periodic structure. For each period of shear the latex particles go back close to their reference position they occupied at time $t_w$. This explains why the correlation fonctions rises back for each period. This shape is usually called an echo\cite{Hebraud97}. However, the thermal and  shear induced activations lead to a certain degree of irreversibility in the scatterers motion \cite{Hebraud97,Holer97,Petekidis02}. This is reflected by a change in the height of the echo. A complete description of the technique can be found in \cite{Hebraud97}. 

Figure \ref{dynaforce}(b) shows that despite the application of the shear the systems keeps on aging over a long period of time. Indeed the height of the echoes keeps evolving even after 1000s.  This shows that a 5.5\% shear is not able to drive the system out of its glassy phase, since no equilibrium could be found over the time range 1s-1000s. We do not claim, that there is no equilibrium state that could be reach after a longer time spent under shear; however, for the time range we studied our system presents a typical aging behavior. We would now like  to evaluate the amplitude above which the system can reach a steady state. A straightforward way to do so would be to increase the amplitude of the shear and look when the dynamics under shear reaches a steady state. This amplitude could then be considered as the limit of the glassy region. However the echo technique fails before such an amplitude can be found. Indeed for too high shear amplitude the echo technic suffers from sampling problems: the camera acquisition rate being too small, the height of the echoes is then measured via one single data point whose value do not necessarily corresponds to the echo's real height. Another scheme has to be found.
    \par
\begin{figure}[th]
\begin{center}
\includegraphics[width=8cm,clip]{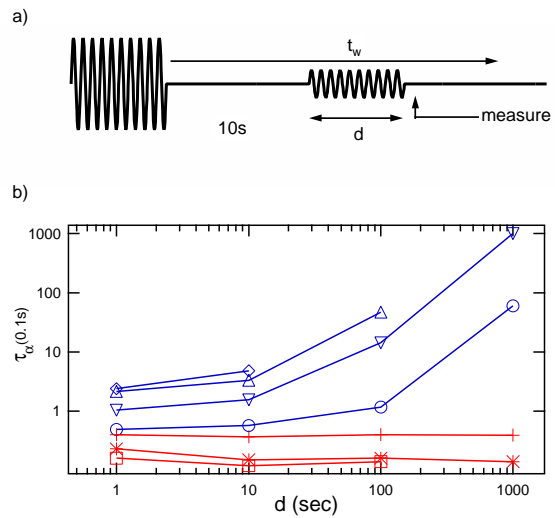}
\end{center}
\caption{a) Sketch of the shear history undergone by the sample. One correlation function is measured 0.1s after the shear cessation and its half decorrelation time $\tau_{\alpha}(0.1s)$ is evaluated.  b)$\tau_{\alpha}(0.1s)$ versus the duration of the second burst for different amplitude $a$. from top to bottom $a=\diamond$ 1.9\%, $\bigtriangleup$ 2.9\%, $\bigtriangledown$ 5.9\%, O 7.9\%, + 8.9\%, $\ast$ 14.9\%, $\square$ 16.9\%. Notice that for $a\geq 8.9\%$ a steady state is reached. The glassy phase boundary lies between 7.9\% and 8.9\%.}
\label{diagphi}
\end{figure}

 We proceed as described on figure \ref{diagphi}(a). The sample is fully rejuvenated by a high shear. On technical grounds the shear is stopped for 10 s after which the oscillation is switched back on at a frequency of 1Hz for different amplitudes $a$ for various durations $d$. The oscillation is then stopped and an intensity correlation function is measured 0.1s after the shear cessation. Its half decorrelation time $\tau_\alpha (0.1s)$ is then evaluated as a fonction of both $a$ and $d$. The results are displayed on figure \ref{diagphi} (b). 

It clearly shows that for amplitudes inferior to 7.9\%, $\tau_\alpha (0.1s)$ does not reach any steady state for the duration we studied. However for amplitudes superior to 8.9\% a steady state is reached and $\tau_\alpha (0.1s)$ does not depend any more on $d$. In addition, $\tau_\alpha (0.1s)$ does dependent anymore on  $a$ for amplitudes superior to 20\% as explained in the first paragraph. The limit of the glassy phase lies thus between 7.9\% and 8.9\%. Actually, for amplitudes lying between these two values the determination of  $\tau_\alpha (0.1s)$ is made difficult by large fluctuations of its values from experiment to experiment.

 \subsection{high shear behavior}\label{hsb}
\begin{figure}[th]
\begin{center}
\includegraphics[width=8cm,clip]{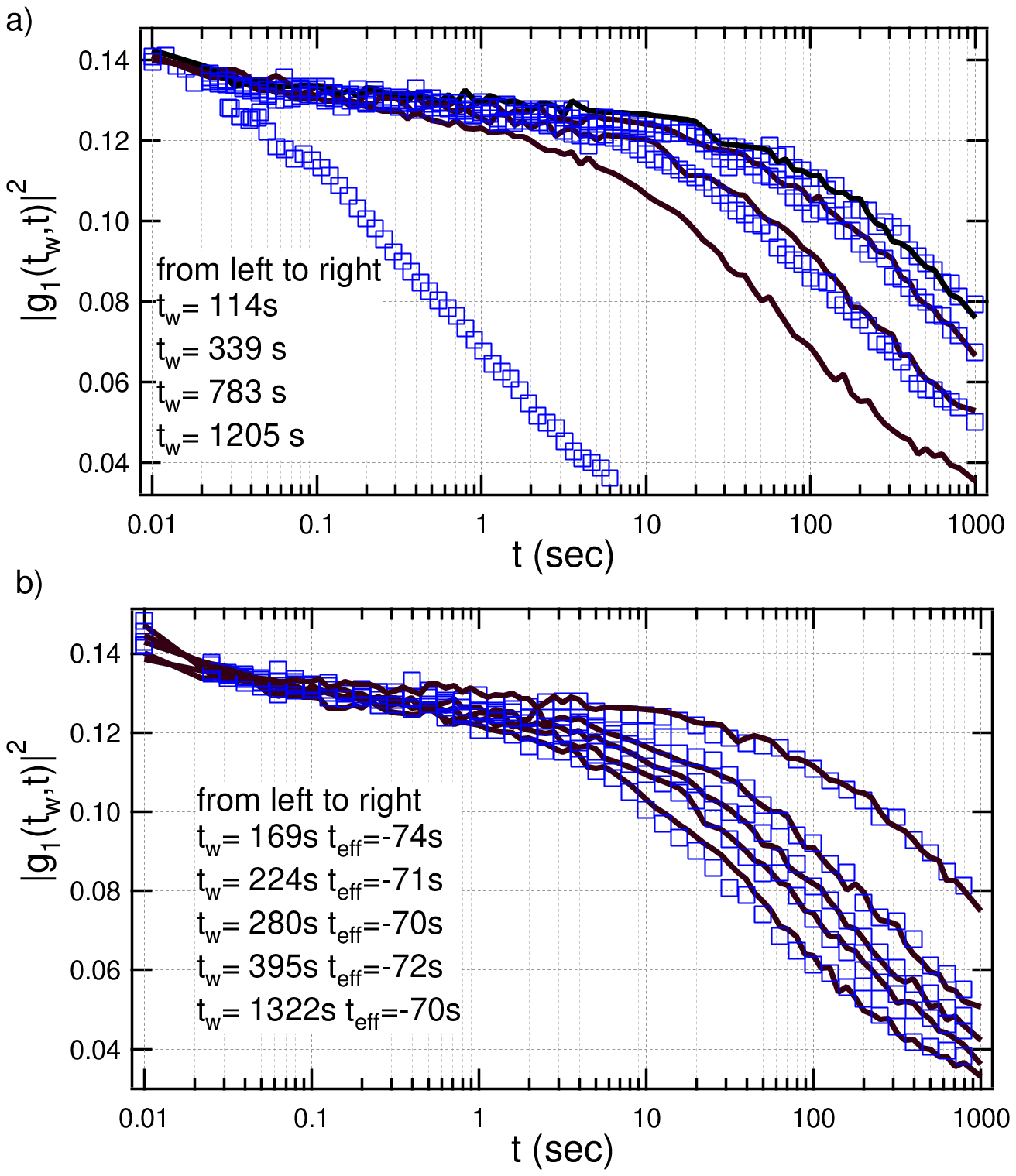}
\end{center}
\caption{a) Same curves as fig \ref{lowshear} but for a shear amplitude of 14.9\%. Notice that \gttwg{14.9} has the same general shape as \gttwg{0}. However the half decorrelation time of \gttwg{14.9} is always \textbf{smaller} than the one of \gttwg{0} for a fixed $t_w$. b) Except for the few very first curves the set of \gttwg{14.9} can be overlayed onto the set of \gttwg{0} by a simple time translation using one single \textbf{negative} $t_{eff}$.}
\label{highshear}
\end{figure}

     We evaluated the boundary of the glassy phase by considering only one point ($\tau_\alpha (0.1s)$) of the first correlation function following the shear cessation. In this section, we now focus on the full set of curves that derive from this shear history. The duration of the second burst is now fixed to 100 seconds. We concentrate on the dynamics following the shear cessation for different amplitudes. As in the previous section the system is fully rejuvenated by a high shear, then it is left at rest for 10s, before it undergoes another burst of oscillation at 1Hz during 100s for an amplitude of 14.9\%. Hence the second burst drives the system out of its glassy region. The set of intensity correlation functions is recorded after the low intensity burst cessation. We call $t_w$ the time elapsed since the cessation of the \textbf{high} shear (this notation is changed compared to previously published papers). We compare the shape of the curves after the shear \gttwg{14.9} to the curves obtained with no second burst: \gttwg{0}. As observed on figure \ref{highshear}(a), it appears that \gttwg{14.9} has the same general shape as \gttwg{0}. However, for the same $t_w$, \gttwg{14.9} decorrelates more quickly than \gttwg{0}. Hence the set of sheared curves seems younger than the unperturbed one. This result is not surprising and is a direct consequence of the fluidisation of the system under shear. The analysis of the curves can be pushed further.\\
Since the global shape of the two sets of curves seems similar it is interesting to see if one can be deduced from the other by a simple shift in time. Thus we computed the value of the effective delay defined as:
$$ |g_1(t_w+t_{eff},t,\gamma=0\%)|^2\simeq\gttwg{14.9}$$ for different $t_w$. For all the $t_w$ that we studied a \textbf{negative} value of $t_{eff}$ could be found that led to a good overlay of the perturbed and unperturbed curves (see figure \ref{highshear}(b)). We found a unique effective delay $t_{eff}=-70s \pm 5s$ for all $t_w$.

Notice also that the perturbed and reference curves at the same age seems to merge on a log scale for large enough $t_w$. This results from the fact the perturbed and unperturbed set of curves can be deduced one from the other by a simple shift in time. At long enough $t_w$, the differences between $t_w$ and $t_w+t_{eff}$ can no longer be seen.

    \subsection{Low shear behavior}\label{lsb}
\begin{figure}[th]
\begin{center}
\includegraphics[width=8cm,clip]{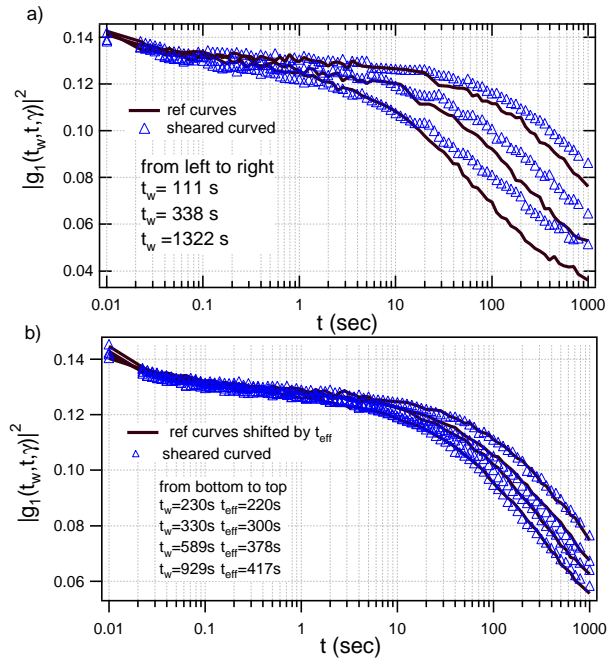}
\end{center}
\caption{a) Comparaison of \gttwg{2.9} and \gttwg{0}, for different values of $t_w$. The perturbed curves have the same global shape as the unperturbed one. However the half decorrelation time of \gttwg{2.9} is always \textbf{larger} than the one of \gttwg{0} for a fixed $t_w$. We called this effect overaging. Notice that $t_w$ is measured from the cessation of the high shear. Hence $t_w-110s$ is the time elapsed since the stop of the second burst. b) Overlay of $|g_1(t_w,t,\gamma=2.9 \%)|^2$ and $|g_1(t_w+t_{eff},t,\gamma=0\%)|^2$. As the legend indicate the overlay is made possible by changing the value of $t_{eff}$ for each $t_w$.}
\label{lowshear}
\end{figure}

    We now focus on the behavior following a small amplitude shear (2.9\%). From figure \ref{diagphi} it appears that the systems keeps on aging under the mechanical oscillation. One could think that the spontaneous aging is nonetheless hindered by the shear. Actually, this is the opposite that happens: the sollicitation helps the system to age as demonstrated on figure \ref{lowshear}(a). We compare the shape of the curves after the shear \gttwg{2.9} to the curves obtained with no second burst: \gttwg{0}. It appears that \gttwg{2.9} decorrelates \textbf{more slowly} than \gttwg{0}. Hence the set of sheared curves seems older than the unperturbed one. 

\begin{figure}[th]
\begin{center}
\includegraphics[width=8cm,clip]{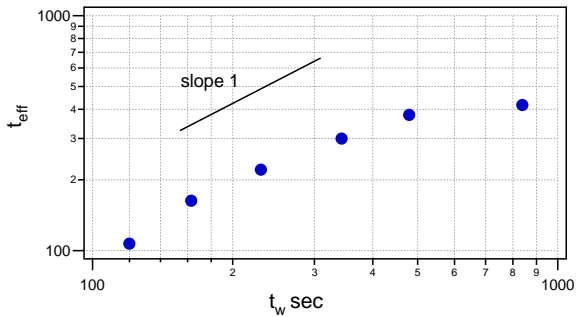}
\end{center}
\caption{$t_{eff}$ versus $t_w$. The effective delay increases with $t_w$ and seems to plateau for large values of $t_w$. The shear amplitude $a$ is 2.9\%.}
\label{teff}
\end{figure}

It clearly shows that the effect of low shear is to age the system faster. We called this effect overaging. Since the global shape of the two sets of curves seems similar it is interesting to see if one can be deduced from the other by a simple shift in time. Thus we computed the value of the effective delay. A \textbf{positive} value of $t_{eff}$ could be found that led to a good overlay of the perturbed and unperturbed curves (see figure \ref{lowshear}(b)) for all studied $t_w$. However the value of $t_{eff}$ now varies with $t_w$ as it is shown on figure \ref{teff}. It thus clearly demonstrates that even though each perturbed curve can be deduced from an unperturbed curve by a simple time translation, the \textbf{full set} of sheared curves cannot be inferred that way from the unperturbed one. 

It thus differs from the high shear case. However, $t_{eff}$ seems to plateau for large waiting times. For long waiting time the ratio $\frac{t_{eff}}{\tau_\alpha}$ tends to zero. Indeed $\tau_\alpha$ increases as $t_w$ whereas $t_{eff}$ tends to plateau. Hence $t_{eff}$ becomes increasingly difficult to determine as $t_w$ is getting larger.  It also  implies that on a log scale the perturbed correlation functions and the unperturbed ones tend to merge. This behavior is thus the same as the behavior for high shear.
    \subsection{Intermediate shear behavior}\label{isb}
\begin{figure}[th]
\begin{center}
\includegraphics[width=8cm,clip]{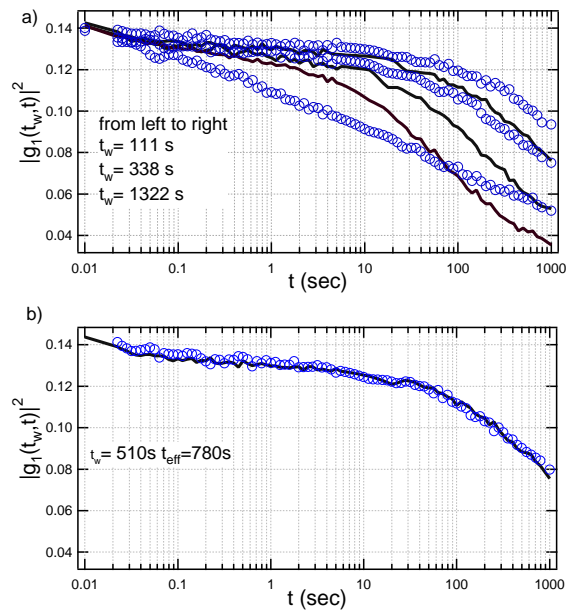}
\end{center}
\caption{a) Same curves as fig \ref{lowshear} but for a shear amplitude of 5.9\%. Notice the crossing of \gttwg{5.9} and \gttwg{0} for small $t_w$. For larger $t_w$ \gttwg{5.9} seems overaged compare to \gttwg{0}. b) In the limit of large $t_w$ \gttwg{5.9} resume its reference shape and can be overlayed to a reference curve via a time translation. In that case $t_{eff}=780s$}
\label{Intershear}
\end{figure}

If now the amplitude of the second burst is set at a higher value, the system behavior differs from the one we have just described. We set the amplitude $a$ of the secondary oscillation to the value of $5.9\%$ without changing the other parameters. As previously explained the set of perturbed curved recorded after the shear cessation is compared to that of the unperturbed case. The results are displayed on figure \ref{Intershear}(a). The main difference from the preceding case lies in the fact that the shape of \gttwg{5.9} is not the same as the one of \gttwg{0}. 

The change in shape can be described as follows: 
\newline - \textbf{For short $t_w$} (1s), the function \gttwg{5.9}, decreases faster than \gttwg{5.9} at short time t but slower at long time t. The dynamics thus seems partially rejuvenated at short t and overaged at long t. It follows that the two curves for a fixed $t_w$ eventually cross one another.
\newline - \textbf{For longer $t_w$} (1212s), the rapid decrease at short time disapears and the function \gttwg{5.9} decorrelates more slowly than the unperturbed one. One could notice that the shape \gttwg{5.9} in the limit of large $t_w$ is close to that of the unperturbed case. Indeed in that limit one can again find a \textbf{positive} effective delay $t_{eff}$ that allows the overlay of an unperturbed curve and a sheared one (fig  \ref{Intershear}(b)). This delay also depends on $t_w$. As in the former case the ratio $\frac{t_{eff}}{t_w}$ seems to tend towards 0  and the perturbed and reference curves merge on a log scale in the limit of large $t_w$.

\section{Discussion}
    In some previous articles\cite{viasnoffprl02,ViasFara02}, we explained how the observed phenomenology can be described by the Soft Glassy Rheology model \cite{Sollich97,Sollich98}. In this section we try to extract some physical ingredients that characterize the change of the microscopic dynamics of our system under shear. We analyze our data with no reference to any specific theoretical model in order to make our conclusions as general as possible. We first show that our results are fully compatible with a weak long term memory (WLTM)\cite{Bouch97}, and that they can be consistently understood considering the full distribution of relaxation times in the system. We then comment upon the ability to  rejuvenate totally the sample with a  shear of finite amplitude applied for a finite duration. We also emphasize the differences between our observed rejuvenation and the phenomenon called rejuvenation for spin glasses. Eventually we propose a spatial scenario that could explain the phenomenology.

\subsection{Weak long term memory}
    In the long waiting time limit the three regimes display a common behavior: the perturbed and unperturbed curves seems to merge, at least on a log scale. In this section we will show that it can be interpreted in terms of Weak Long term Memory \cite{Bouch97}.\\
In a simple aging experiment, the correlation fonctions \gttwg{0} rescale onto a master curve. It thus shows that their evolution during such a type of experiment is autosimilar in time: \gttwg{0} can be written as a simple function $f(\frac{t}{t_w})$ of the variable $\frac{t}{t_w}$. However, it is not straightforward that any shear history would lead to correlation functions that can be written that way. Nonetheless, the experimental section shows that in the long time limit the perturbed correlation functions $|g_1(t_w,t,\gamma)|^2$ approximately write as $|g_1(t_w+t_{eff},t,\gamma=0\%)|^2$. In the case of a large amplitude, one can define a constant negative effective delay, in the case of a small sollicitation $t_{eff}$ varies with $t_w$, however $t_{eff}$ seems to plateau at large $t_w$. For intermediate $\gamma$ the situation is slightly different; no $t_{eff}$ can be found for most values of $t_w$. The shapes of the correlation fonctions clearly differ from the shape of the simple aging one. However, in the limit of large $t_w$ the shape of the perturbed curves tend towards the reference one. Hence for any strain history and  in limit of large $t_w$, the perturbed correlation functions can be written as:
$$|g_1(t_w,t,\gamma)|^2\simeq|g_1(t_w[1+\frac{t_{eff}}{t_w}],t,\gamma=0\%)|^2$$
which also reads, using the master function $f$:
$$|g_1(t_w,t,\gamma)|^2\simeq f\left(\frac{t}{t_w+t_{eff}}\right)\simeq f\left(\frac{t}{t_w+o(t_w)}\right)$$
It physically means that, whatever its initial dynamical state, the system will always spontaneously evolves towards a unique asymptotic dynamical state. The effect of any finite perturbation applied for a finite amount of time will thus eventually fade away.\\
 This observation is reminiscent of the concept of weak long term memory (WLTM)\cite{Bouch97}. WLTM means that the system is able to explore its whole phase space whatever its initial state and even though some regions may take a very long time to be reached. As a consequence, the disturbance of the dynamics by a finite perturbation will disappear in the limit of large waiting times. Even though this principle has been rigourously observed in many theoretical models, it is hardly discussed in experiments. Our results are good hints that such a principle applies to soft glassy materials, at least from the point of view of their microscopic dynamics. 

\subsection{Modifications of the relaxation time distribution}
    In the preceding sections, we discussed the shape of the correlation fonctions. An equivalent way of describing them is to consider their distributions $P(\tau_{dec})$ of decorrelation times. Theses distributions are linked to the distributions of relaxation times $P(\tau_{rel})$ in the sample. This point of view is able to give us a coherent image of the observed phenomenology.
    \par The aging process viewed from the point of view of \dr corresponds to a slow drift of the distribution towards slower and slower relaxation times. Furthermore, the rescaling of  all the correlation functions implies that this slow drift happens in a self similar manner. Hence, the distribution must rescale with the waiting time. It thus must write:
$$P(\tau_{rel},t_w)=P_m(\frac{\tau_{rel}}{t_w})$$
where $P_m$ is the master function for P. Similarly, the weak long term memory behavior can be understood as the convergence of any initial distribution towards a unique asymptotique master function $P_m$ in the limit of large waiting times. At that point, the observed phenomenology can be understood as follows:\\
\begin{itemize}
\item When the perturbation drives the system out of its glassy phase, the distribution is shifted towards faster relation times. It is, within the experimental precisions, similar to a distribution through which the system passes during its reference evolution. In fact a simple time translation allows a good overlay of the whole set of correlation curves.
  \item Oppositely, a small amplitude perturbation leads to an overaging of the system. It means that the microscopic dynamics is slowed down after the perturbation. As a result the distribution is shifted towards longer relaxation time. This result is not intuitive, and the nature of the shift is not trivial. One could think that it corresponds to a speeding up of the autosimilar drift of the distribution. If it were so, the full set of the perturbed correlation functions would corresponds to a time translation of the reference one via a \textbf{unique} effective delay $t_{eff}$. However, we demonstrated, in section \ref{lsb}, that $t_{eff}$ actually depends on $t_w$. It implies that the perturbation enriched the distribution P with slow relaxation time in a way that differs from that of a spontaneous evolution. This point will be further discussed in section \ref{relteff}.
  
  \item For intermediate amplitudes, the shape of the distribution P is greatly modified compared to the reference case. This modification entails the drastic change in shape of the correlation function. The distribution is enriched simultaneously with short and long relaxation times. This dynamical state is reflected by the faster decrease of $|g_1(t_w,t,\gamma)|^2$ for small $t$ and its slower decrease for large $t$. This holds for time $t_w$ shortly following the shear cessation. However, for larger $t_w$, we have already mentioned that the system shows a weak long term memory. It means that the dynamical state that we have just described eventually converges towards the distribution $P_m$. Our results further show that this convergence occurs in the following fashion: the short time part of P reaches quickly retrieved its unperturbed shape. It is reflected by the rapid disappearance of the fast decrease of the perturbed correlation function for small $t$. The large relaxation times, however, remain overpopulated compared to the reference history. The correlation function thus show a slowest decay at large $t$. The rejuvenated part of the distribution has disappeared, and the overaged part is left alone.  A positive effective delay can thus be found. For theses values of $t_w$, the intermediate and small amplitude cases are equivalent.
\end{itemize}
Theses conclusions and interpretations are illustrated on figure \ref{overlaydistrib}(left) and do not depend on any underlying model. However, in a previous paper we showed that such a phenomenological scenario can be achieved within the frame of the Soft Glassy Rheology Model\cite{ViasFara02}.

    \subsection{relevance of the effective delay description}\label{relteff}
    The literature of glasses often describes the effect of a small positive or negative shift in temperature in term of effective age or effective delay. The two usual ways to evaluate the effective delay for these systems consist in either measuring one time quantities such as the susceptibility or the dielectric constant at a fixed frequency $(\chi"(\omega),\varepsilon"(\omega))$ or two times quantities such as relaxation functions $(M(t_w,t),G(t_w,t))$. The effective delay is determined by shifting the curves after the perturbation onto a reference curve. Apart from some deviations shortly after the perturbation cessation, a good overlay can be found. In this section we emphasize the fact that some misinterpretations can arise if the full evolution of a two times quantity is not considered to evaluate the effective delay.
\par
\begin{figure}[th]
\begin{center}
\includegraphics[width=8cm,clip]{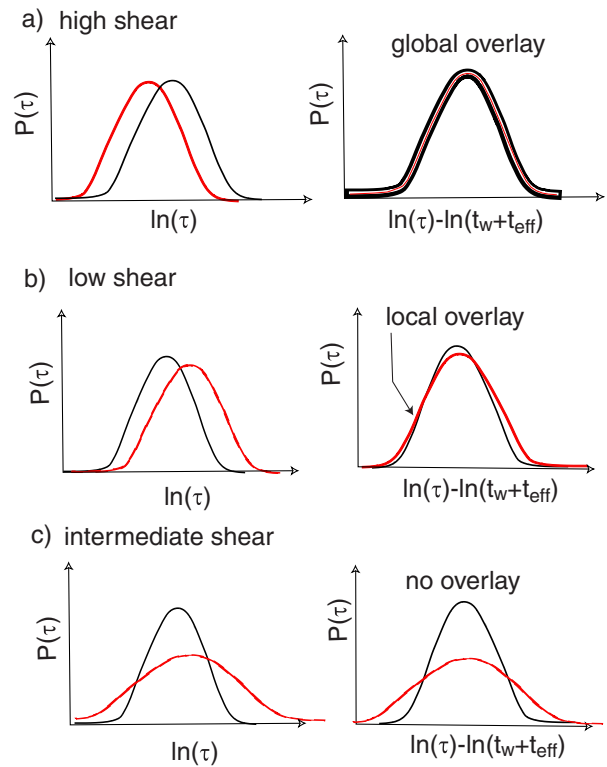}
\end{center}
\caption{Sketch of the different distribution shape leading to an effective delay. Left: reference distribution (black) and perturbed distribution (gray) Right: overlay of the two distributions. a) Case of a autosimilar shift that leads to a perfect overlay over the full range of the distribution of relaxation time. In that $t_{eff}$ is independent of $t_w$.
b) Case of the overaging process where the shape of the distribution is modified by the shear application. The change in shape is such that a local overlay can be achieved over the tested part of the relaxation time. The overlay being local and not global, $t_{eff}$ varies with $t_w$. c) Case where the change in shape does not allow any local overlay }
\label{overlaydistrib}
\end{figure}

 If the description of the change in the dynamics in terms of effective delay would be an exact result, it would then mean that the perturbation drives the system into a dynamical state through which it would have passed during a simple aging history. Hence an effective age would be fully sufficient to characterize this state. In this event, the relaxation time distribution after the perturbation at time $t_w$ would be truly identical to the reference distribution at time $t_w+t_{eff}$. This is schematically presented on fig \ref{overlaydistrib} a). As a consequence the evolution of the perturbed distribution would be the exact evolution of the reference one but shifted by the same delay $t_{eff}$.

In this event the determination of an effective delay by the overlay of one single correlation curve with a reference one, or the overlay of the evolution of a single typical point of each correlation function with the evolution of that same point for a reference history, would lead to the same result. This is what happens in the high shear regime (see fig \ref{highshear}) where $t_{eff}$ is found constant with $t_w$.
However, in the low shear regime, $t_{eff}$ depends on $t_w$.  
This implies that looking at the evolution of a single point value is thus no longer equivalent to looking at the system at a fixed $t_w$. The shape of the distribution is changed. However, if the change is slight for relaxation times corresponding to \textbf{the time window over which the measurement is performed}, an apparent $t_{eff}$ can be found.
However, if the evolution of the distribution depends on the \textbf{entire spectrum of relaxation times}, the evolution of the probed part of the distribution may differ from the reference one. This situation is illustrated on figure \ref{overlaydistrib} b). Although the distribution seems to overlay over a certain time range a t a fixed time, their evolution differs, since the evolution takes into account the whole distribution.
Hence the value $t_{eff}$ has to be varied with $t_w$ in order to keep a good overlay of the reference and perturbed distributions. \\ 
We remark that this variation in time of $t_{eff}$ cannot be detected by the usual methods described in the first lines of this section. This effect can be measured here since our experiment gives easy access to the \textbf{evolution} of a \textbf{two times quantity}:\gttwc. This is the reason why we think this notion of effective delay should be considered carefully. 
   
\subsection{High shear rejuvenation}\label{hsr}
    Macroscopic observations clearly shows that the rheological properties of such systems can reach a steady state under high shear\cite{Derec00,Cloitre00,Bonn02,VanMeg94}. The shear cessation acts then as a thermal quench from a high temperature state. However, the type of shear that are usually used are either continuous ones at a fixed shear rate, or oscillatory ones at very high amplitude (typically 1000\%). We now show that, in the case of a sample with a volume fraction close to that of its dynamical transition, the microscopic dynamics can be entirely renewed applying an oscillatory shear of amplitude around 20\%. This shear sollicitation is able accelerate the rate of the structural relaxations down to the shortest time that our set up can measure $\sim 0.05s$.
A careful look at the correlation functions displayed on figure \ref{vieilsimple} shows a fast decay of all the correlation functions in their very first points. This decay corresponds to the tail of the fast $\beta$ decay existing in all glasses. They correspond to individual fast relaxation modes that are reminiscent of the relaxation modes in the liquid phase.  The long time decay that we studied here is often referred as the $\alpha$ modes that arise from cooperative structural rearrangements. Theses modes are the signature of the vicinity of the glass transition. Notice that on the correlation function taken  0.1s after the shear cessation, the $\alpha$ and $\beta$ modes are approximately separated by only 1 decade. It thus means that under high shear theses modes are hardly separated in time. The relaxation processes are thus very close to that of a fluid. This is the logical reason why higher or longer shear periods do not modify the dynamics following the shear cessation.
    \par
\begin{figure}[th]
\begin{center}
\includegraphics[width=8cm,clip]{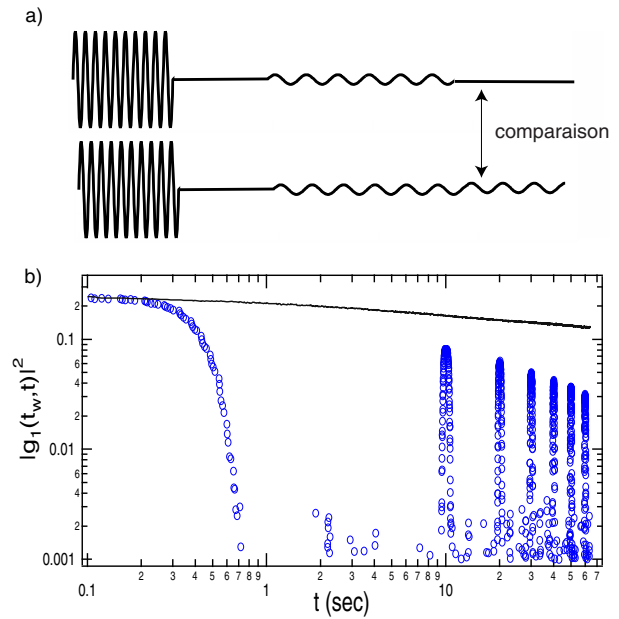}
\end{center}
\caption{a)Shear history for comparison b) Line: reference curve for spontaneous decorrelation after a cycle of 10 oscillation at 0.1Hz at an amplitude of 5.5\%. Symbols: Correlation fonction of the system under shear at 0.1Hz at an amplitude of 5.5\%. Notice that the echoes' height decreases faster than the reference curve.}
\label{shearrest}
\end{figure}

 We also want to point out a real difference between the rejuvenation phenomenon we observe here and the rejuvenation described for spin glasses or polymers. As demonstrated by figure \ref{diagphi}, rejuvenation occurs if the shear amplitude $a$ is superior to 8.9\%. The condition for our sample to be rejuvenated is that it is rheologically quenched from a steady state into a glassy state.  In spin glasses the observed rejuvenation (restarting of the dynamics after a negative temperature step) occurs when the system is thermally quenched from a glassy state into another glassy state. In those systems, the dynamics at lower temperature does no depend on the time spent at a higher temperature. These effects can be described among other descriptions by a hierarchical organisation of the phase space \cite{Hammann92}. In our case, no such descriptions are needed to explain our results since rejuvenation occurs only if the colloidal glass is driven out of its glassy region.\\
In order to prove that there is no rejuvenation effects in the sens of spin glasses, we performed the following experiment. We compare the correlation function of a system \textbf{under shear} to the correlation fonction of the same system left at rest. The shear amplitude is fixed to 5.9\% such that it lies in the region where the system keeps on aging despite the shear application see fig \ref{diagphi}. The shear history is displayed on figure \ref{shearrest} (a). 

Figure \ref{shearrest} (b) shows that the height of the echoes decays faster than the reference curve. Hence, the forced dynamics results in an additional degree of irreversibility in the scatterers motion. It clearly appears that the correlation function decreases faster under shear than when left at rest. Though this experiments is the exact analogous to the negative temperature shift experiment, its results are clearly at odd with the rejuvenation effect for spin glasses \cite{ Vincent96}. We thus conclude that the similarity between temperature shift and oscillating shear only lies in the fact that they transiently modify the rearrangement rate.
\par We finally make the following remark:  In any case (rejuvenation, overaging or crossover regime) the microscopic dynamics is accelerated \textbf{while the system is under shear}. This is fully consistent with the shear thinning behavior of such systems. The different behaviors described in the previous sections result from the spontaneous evolution of the structured formed when the shear is stopped. In that prospect, overaging is a result of the shear thinning behavior \\ 
 On the other hand shear thickening material may present a different phenomenology: the dynamics under shear may appear slower than the spontaneous one. Thus the shear cessation can be followed by an acceleration of the dynamics. Even though such results may look phenomenologically identical to the spin glass rejuvenation effect, they may arise for totally different reasons such as anisotropy and not because of hierarchical organisation of relaxation times. 

        \subsection{Real space proposed scenario}
    In the preceding sections we described our results in terms of distribution of relaxation times. However, we do not know the very nature of the relaxation modes associated with theses times. In this section, we propose a picture in real space of a possible scenario that can occur in such materials. We are aware that this scenario is certainly to simplistic, and can be view as a basis for more elaborated ones.\\
On macroscopic length scales, the sample is affinely deformed by the macroscopic shear. However, the heterogeneous structure of the rheological properties of the material on smaller length scale can result in stress and/or strain concentration on certain regions. If the stress is locally superior to the local yield stress, a local rearrangement is induced by the macroscopic deformation.\\
 $\bullet$ At small shear amplitude, one could think that these rearrangements events occur in the most fragile regions that are randomly distributed in the sample. They may involve few particles. The external shear thus acts as a "boost" for the rearrangement rate. and increased probability is provided to the system to find more stable configuration. The effect of mechanical deformation may thus be to add to the thermal noise a random mechanical noise due to the increased number of rearrangements.\\
$\bullet$ At high amplitudes the shear strain is sufficient to disrupt all structures in the sample. Shear is high enough to overcome all local yield stress. The system is fluidized, an equilibrium dynamics is found that resemble that of a liquid (see fig \ref{diagphi}(b)). This corresponds to the usual scenario of shear thinning and total rejuvenation in such systems.\\
$\bullet$ At intermediate amplitude one can naively imagine that two types of zones coexist: the shear amplitude is such that shear induced rearrangements occur in localized mesoscopic zones of the sample. Though on macroscopic length scale the system seems to be homogeneously shear (no macroscopic shear banding), the strain is localized in fluidized areas that thus have a rapid dynamic under and after shear cessation. Theses rapidly rearranging zones may delimit stiff and thus slow dynamics areas that undergo a small straining and mechanical noise from their boundaries. The small perturbation applied onto theses zones may thus lead to an overaging of their dynamics as in the small shear amplitude scenario. This is would explain the simultaneous appearance of the two trends (rejuvenation and overaging) within the same sample. The volume fraction of the overaged zones would highly depend on the shear amplitude, ranging from 0 for a high amplitude to a macroscopic size for very small amplitudes.\\
We also want to point out that such a scenario occurs only if no macroscopic shear banding occur. In that case a total rejuvenation of the sample by  a stationary shear is not possible. This is why theses effect are only observed in a narrow range of volume fraction as discussed in \cite{ ViasFara02}.

\section{conclusion}
In this article we discussed in detail the effect of shear upon the microscopic dynamics of a colloidal glass. We showed theses effect can be understood as a change in the relaxation time distribution.  Small amplitude shear result in drift of the distribution towards longer relaxation time. This drift happen in a different fashion than the spontaneous aging one. Intermediate amplitude broadens the distribution both in the short and long relaxation time direction. High amplitude shear allow a steady state to be reached that entirely rejuvenates the system. We further show that no matter the initial distribution shape, it always converges towards a unique distribution for long waiting times. We also made a point out of the ambiguity of the determination of an effective age for glassy systems submitted to small perturbations.
\par The authors thank J.P Bouchaud and L Berthier for enlightening discussions.

\end{document}